\documentstyle[psfig,amssymb]{mn}
\title[The Phoenix Deep Survey: The star-formation rates and the stellar
masses of EROs] 
{The Phoenix Deep Survey: The star-formation rates and the stellar masses of 
EROs} 

\author[A. Georgakakis et al.] {A. Georgakakis$^{1}$\thanks{email:
  a.georgakakis@imperial.ac.uk}, A. M. Hopkins$^{2}$, J. Afonso$^{3}$,
  M. Sullivan$^{4}$, B. Mobasher$^{5}$, \\\\
  {\LARGE L. E. Cram$^{6}$}\\ \\  
  $^1$Imperial College of Science Technology and Medicine, Blackett
  Laboratory, Prince Consort Rd, SW7 2BZ, London, UK\\ 
  $^2$School of Physics, Bldg A29, University of Sydney, NSW 2006, Australia\\
  $^3$Centro de Astronomia da Universidade de Lisboa, Observat\'orio
  Astron\'omico de Lisboa, 1349-018 Lisboa, Portugal\\  
  $^4$Department of Astronomy and Astrophysics, University of Toronto,
  60 St. George Street, Toronto, ON M5S 3H8, Canada\\
  $^5$Space Telescope Science Institute, 3700 San Martin Drive,
  Baltimore, MD 21218, USA\\
  $^6$The Australian National University,  Canberra ACT 0200,
  Australia  
}
\begin{document}
\maketitle  

\begin{abstract}
We estimate the star-formation rates and the stellar masses of the 
Extremely Red Objects (EROs) detected in a  $\approx180 \rm \,
arcmin^2$ $Ks$-band survey ($Ks\approx20$\,mag). This sample is
complemented by sensitive 1.4\,GHz radio observations (12$\mu$Jy
 $1\sigma$ rms) and multiwaveband photometric data ($UBVRIJ$) as part
of the Phoenix Deep Survey. For bright $K<19.5$\,mag  EROs in this
sample ($I-K>4$\,mag; total of 177) we use photometric methods to
discriminate dust-enshrouded active systems from  early-type
galaxies and to constrain their redshifts. Radio stacking is then
employed to estimate mean radio flux densities of $\approx 8.6$
($3\sigma$) and $6.4\,\mu$Jy ($2.4\sigma$) for the dusty and
early-type subsamples respectively. Assuming that dust enshrouded
active EROs are powered by star-formation the above radio flux
density at the median redshift of $z=1$ translates to a radio
luminosity of $L_{1.4} = 4.5 \times  10^{22} \rm W/Hz$  and a
star-formation rate of  $\rm SFR = 25 \, M_{\odot} \,
yr^{-1}$. Combining this result with photometric redshift estimates we 
find a lower limit to the star-formation rate density of $0.02\pm0.01
\rm \, M_{\odot}  \, yr^{-1} \, Mpc^{-3}$ for the $K<19.5$\,mag dusty
EROs in the range $z=0.85-1.35$. Comparison with the star-formation  
rate density estimated for previous ERO samples (with similar
selection criteria) using optical emission lines, suffering dust
attenuation, suggests a mean dust reddening of at least
$E(B-V)\approx0.5$ for this population.  We further use the $Ks$-band
luminosity as proxy to stellar mass and argue that the dust enshrouded
starburst EROs in our sample are massive systems, $\rm M \ga 5 \times
10^{10} \, M_{\odot}$. We also find that EROs represent a sizable
fraction (about 50 per cent) of the number density of galaxies more
massive than $\rm M = 5 \times 10^{10} \, M_{\odot}$ at $z\approx1$,
with almost equal contributions from dusty and early type
systems. Similarly, we find that EROs contribute about half of the
mass density of the Universe at $z\approx1$ (with almost equal
contributions from dusty and early types), after taking into account
incompleteness because of the magnitude limit $K=19.5$\,mag.      
\end{abstract}


\section{Introduction}\label{sec_intro}
The class of Extremely Red Objects (EROs; $R-K>5$, $I-K>4$\,mag),
first identified more than 15 years ago (Elston  et al. 1988), is
believed to comprise a heterogeneous population of $z \ga 1$ systems
split between passive galaxies and dust enshrouded AGNs/starbursts
(Cimatti et al. 2003). The identification of either type of galaxies
(early or dusty) at high-$z$ has important  cosmological implications,
thus providing significant impetus in  ERO studies.   

For example, finding early-type massive systems at high-$z$ can
provide information on both the galaxy formation redshift (Spinrad et
al. 1997; Cimatti et al. 2002) and the global mass assembly
(e.g. Fontana et al. 2003, 2004; Drory 2004; Glazebrook 2004), thus
constraining galaxy formation scenarios: Monolithic collapse early in
the Universe ($z_f>2-3$) followed by passive evolution (e.g. Eggen et
al. 1962; Larson 1975) versus hierarchical merging and relatively
recent formation epochs (Baugh et al. 1996; Kauffmann 1996). Similarly
a population of $z \ga 1$ dusty active galaxies, AGNs or
starbursts, that are missing from UV/optical surveys may play an
important role  in the global star-formation history (e.g. Haarsma et
al. 2000; Smail et al. 2002; Hopkins 2004), the evolution of AGNs and
the interpretation of the diffuse X-ray Background with respect to the
(still) elusive population of high-$z$ obscured QSOs (e.g. Hasinger et
al. 2003).   

Recent developments in instrumentation have yielded large ERO samples
allowing  systematic study of their statistical properties in a
cosmological context. Cimatti et al. (2002) used optical spectroscopy
from the {\sf K20} survey to explore the star-formation rates (SFR) of
$K<19.2$\,mag EROs and to assess their contribution to the  global
star-formation history. Under conservative assumptions about
reddening, these authors find that  dust-enshrouded EROs represent a
small but non-negligible fraction (about 10 per cent) of the SFR
density at $z\approx1$. Dust obscuration issues, however, make this
result uncertain. Smail et al. (2002) expanded on the Cimatti et
al. (2002) study using deep radio imaging to explore the SFR of
$K<20.5$\,mag EROs independent of dust induced biases. Using this 
deeper $K$-band sample they estimate star-formation densities higher
than those of Cimatti et al. (2002) and suggest that obscured galaxies
make a sizable contribution to the total SFR density at
$z\approx1$. More recently 
Caputi et al. (2005) used the GOODS-South data to investigate the
evolution of $K$-band selected galaxies. They argue that EROs among
their sample constitute a sizable fraction (about 50-70 per cent) of
galaxies with stellar mass $\rm M>5\times 10^{10}\,M_{\odot}$ at
$z=1-2$, suggesting that they represent a major component of the
stellar mass build-up at these redshifts.     

The observational developments above are also complemented with
efforts to model EROs using either semi-analytical (e.g. Somerville et
al. 2004a) or hydrodynamical (e.g. Nagamine et al. 2005)
methods. Despite significant progress, accounting for both the red
colours and the number density of EROs remains a challenge for these
numerical simulations. Part of the difficulty lies in our poor
understanding of some of the properties of EROs. Open questions
include what are the dust properties of these systems, what is the
number density of dusty and early type EROs, what is the relative
contribution of these sub-populations to the mass density. 

In this
paper we add to the discussion on the cosmological significance of
EROs by combing an $\approx 180 \, \rm arcmin^{2}$ deep
($Ks\approx20$\,mag) $Ks$-band survey with ultra-deep 1.4\,GHz radio
data ($\approx \rm 60 \, \mu Jy$) carried out as part of the Phoenix
Deep Survey (Hopkins et al. 2003). This sample has already been
used to explore the clustering properties and the environment of EROs
(Georgakakis et al. 2005). The advantage of our survey is depth
combined with wide area coverage reducing cosmic variance
issues. Additionally, the ultra-deep radio data allow SFR estimates
for EROs free from the dust obscuration effects that are expected to
be important in this class of sources. Throughout the paper we adopt
$\rm H_{o}=70\,km\,s^{-1}\,Mpc^{-1}$, $\rm \Omega_{M}=0.3$ and  $\rm
\Omega_{\Lambda}=0.7$.    

\section{The Phoenix Deep Survey}\label{phoenix}

The Phoenix Deep Survey (PDS\footnote{\sf
http://www.atnf.csiro.au/people/ahopkins/phoenix/}) is an on-going 
survey studying the nature and the evolution of sub-mJy and $\rm
\mu$Jy radio galaxies. Full details of the existing radio, optical and
near-infrared (NIR) data can be found in Hopkins et al. (2003), Sullivan
et al. (2004) and Georgakakis et al. (2005); here we summarise the
salient details. The radio observations were carried out at the
Australia Telescope Compact Array (ATCA) at 1.4\,GHz during several
campaigns between 1994 and 2001, covering a 4.56 square degree area
centered at RA(J2000)=$01^{\rm   h}11^{\rm m}13^{\rm s}$
Dec.(J2000)=$-45\degr45\arcmin00\arcsec$. A detailed description of
the radio observations, data reduction and source detection are
discussed by Hopkins et al. (1998, 1999, 2003). The observational
strategy adopted resulted in a radio image that is homogeneous within
the central $\rm \approx1\,deg$ radius, with the $1\sigma$ rms noise
increasing from $\rm 12\mu Jy$ in the most sensitive region to about
$\rm 90\mu Jy$ close to the edge of the $\rm 4.56\,deg^2$ field. The
radio source catalogue consists of a total of 2148 radio sources to a
limit of 60\,$\rm \mu$Jy (Hopkins et al. 2003).

The $Ks$-band NIR data of the central region of the PDS were 
obtained using the SofI infrared instrument at the 3.6\,m ESO New
Technology Telescope (NTT).  The observational strategy and details of
the data reduction, calibration and source detection are described by
Sullivan et al. (2004). The $Ks$-band mosaic covers a $\rm 13.5 \times
13.3 \, arcmin^2$ area with a 45\,min integration time, and a central
$\rm 4.5 \times 4.5 \, arcmin^2$ subregion which has an effective
exposure time of 3\,h. The completeness limit is estimated to be
$Ks\approx20$\,mag for the full mosaic and $Ks\approx20.5$\,mag for
the deeper central subregion.

Additional $J$-band NIR data of the central region of the
PDS have been obtained using the InfraRed Imaging Spectrograph 2 (IRIS2)
mounted at the f/8 focus of the 3.9\,m Anglo-Australian telescope. These
observations were carried out in 2003 September 8 in photometric
conditions. The IRIS2 is equipped with a 1024x1024 HgCdTe array giving
at the f/8 focus a pixel scale of 0.45\,arcsec and a field of view of
$7.7^{\prime} \times 7.7^{\prime}$. For the observational strategy a
dithering pattern  was adopted that consisted of a sequence of 60\,s
integrations followed by an offset of the telescope (maximum shift
75\,arcsec). To get accurate sky frames the offset vector was not
replicated between successive exposures. A single pointing was
obtained with a total integration time of 1.8\,hours.  

The data reduction was carried out using {\sc iraf} tasks. The flat  
field frame was constructed using both the target observations and the
dome flat following the method developed by Peter Witchalls and Will 
Saunders and described by Sullivan et al. (2004). The individual
sky-subtracted images of the pointing were then combined to produce
the final mosaic which has a useful area, after clipping noisy regions
close to the field edge, of about $\rm 8 \times 8 \,
arcmin^2$. Photometric calibration was carried out using standard 
stars from Persson et al. (1998). The photometric solution
has an rms scatter of about
0.02\,mag. Astrometric calibration was performed using  the positions
of about 40 stars from the SUPERCOSMOS catalogue. This solution has an
rms scatter of about 0.2\,arcsec.  The completeness limit is estimated
to be $J\approx21.5$\,mag. 

Deep multicolour imaging ($UBVRI$) of the PDF has been obtained using
the Wide Field Imager at the AAT ($BVRI$-bands) and the Mosaic-II
camera on the CTIO-4m telescope ($U$-band), fully overlapping the SofI
$Ks$-band survey. Full details on the data reduction, calibration and
source detection are again presented in Sullivan et al. (2004). In
this study we will use the $I$-band observations, these being the
deepest ($I\approx24.2$\,mag; see next section) and most appropriate
for identifying EROs.

\section{The ERO sample and photometric redshifts}\label{sec_photoz}
The ERO sample selection is described in detail by Georgakakis et
al. (2005). In brief  EROs are selected to have $I-K\ge4$\,mag. The
$5\sigma$ detection threshold for the $I$-band catalogue is 24.2\,mag, 
sufficiently deep to identify  EROs with $Ks=20$\,mag. We find a
total of 289 EROs to this magnitude limit within the $\rm 13.5 \times
13.3 \, arcmin^2$ area of our survey. To avoid incompleteness at faint
$Ks$-band magnitudes and low signal-to-noise ratio data, in the
analysis that follows we consider only those EROs with $K<19.5$\,mag,
giving a total of 177 sources.  

In the absence of optical spectroscopy we attempt to classify the ERO
sample into different types, dusty and evolved, using their
multiwaveband photometric properties. We adopt a method similar to
that described by Smail et al. (2002) to fit two different template
SED families (``dusty" or ``old") to the optical/NIR magnitudes of EROs. We
use the {\sc hyper-z} code (Bolzonella, Miralles \& Pell\'o
2000) to estimate photometric redshifts and to assess the goodness of
fit for the two set of models. We then adopt the model and the
corresponding photometric redshift that best fits the observations
(minimum $\chi^2$). In the case of dusty systems we adopt a continuous  
star-formation model with Salpeter IMF and solar metallicity as
implemented in the new isochrone synthesis code  of Bruzual \& Charlot
(2003). The reddening is allowed to vary in the 
range $\rm A_V=1-4$ (e.g. Cimatti et al. 2002) assuming the extinction
curve of Calzetti et al. (2000). For evolved galaxies we use a model
SED with exponentially declining star-formation rate with e-folding time
of 1\,Gyr, Salpeter IMF and solar metallicity (Bruzual \&
Charlot 2003). Only mild reddening is allowed for this set of template
SEDs with $\rm A_V<0.5$. To minimise spurious redshift estimates we
consider only those EROs with $K<19.5$\,mag and at least 3 band 
detections. The latter excludes from the sample a total of 40 
sources, i.e. about 20 per cent of the $K<19.5$\,mag sample. Each ERO
is assigned the template (dusty or old) and the corresponding
photometric redshift that gives the minimum $\chi^2$. For a small
number of sources (14) neither dusty nor old SEDs can provide an
acceptable fit, both giving $\chi^2>2.7$, i.e. rejection of the model
templates at the $>90$ per cent confidence level. These 14 systems are
excluded from the analysis. We note that these sources either have more
complex SEDs that those assumed here, are associated with QSO
dominated sources or are Galactic stars, e.g. late type cool dwarfs
(e.g. Francis et al. 2004). For example 4 of the 14 sources are
assigned {\sc SExtractor} flag {\sc class\_star}$=0.93-0.95$ only
marginally lower than the cutoff used for star/galaxy separation ({\sc
class\_star}$>0.95$; Georgakakis et al. 2005). The final sample
comprises a total of 123 $K<19.5$\,mag EROs.   

The method above clearly, cannot reproduce complex SEDs with
contributions from both evolved and young obscured stellar
populations. Nevertheless, it provides a rough classification allowing
an assessment of the relative fraction of EROs at the two extremes:
Old and passively evolving vs young, reddened systems. Some fraction
of misclassifications are however, inevitable in the  scheme  
above that relies on  broad-band optical/NIR photometry to
differentiate between dusty and old EROs. Moreover, some of the EROs 
in the sample do not have $J$-band photometry available, which is
sensitive to the H+K spectral break of evolved galaxies at  $z \ga 1$.
This is likely to introduce further uncertainty into the
classification. Also, a fraction of EROs are expected to harbor
type-II AGN activity. Here we assume that it is the host galaxy
stellar population, rather than the obscured AGN that dominates the
broad-band optical/NIR colours of these systems. Observations of X-ray
selected obscured AGNs both at moderate and higher-$z$ suggest that
their continuum emission is indeed dominated by the host galaxy rather
than the central engine  (e.g. Gandhi et al. 2004; Georgakakis et
al. 2004; Mobasher et al. 2004).   

\begin{figure}
 \centerline{\psfig{figure=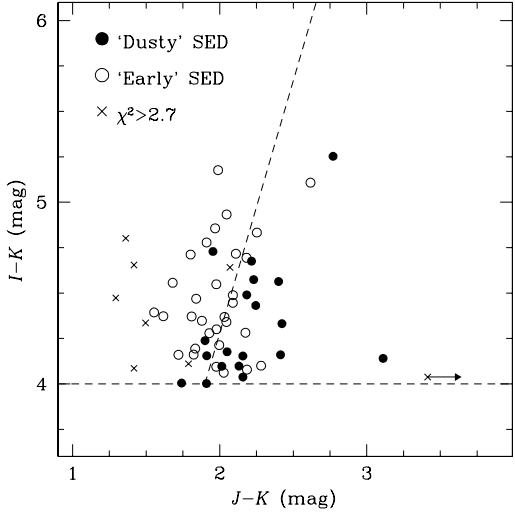,width=3in,angle=0}}
\caption
 { $I-K$ against $J-K$ colour plot. The horizontal line shows our ERO
 selection $I-K>4$\,mag. The diagonal dotted line, introduced by  Mannucci
 \& Pozzetti (2001), discriminates between early (on the left) and
 dusty (to the right) systems. The filled and open circles are
 $K<19.5$\,mag EROs in the PDS that are best fit by dusty and early
 template SEDs respectively. Crosses correspond to EROs for which the
 template fitting method described in the text does not provide an
 acceptable fit to the observations, giving $\chi^2>2.7$.
 }
\label{fig_colour}
\end{figure}

\begin{figure}
 \centerline{\psfig{figure=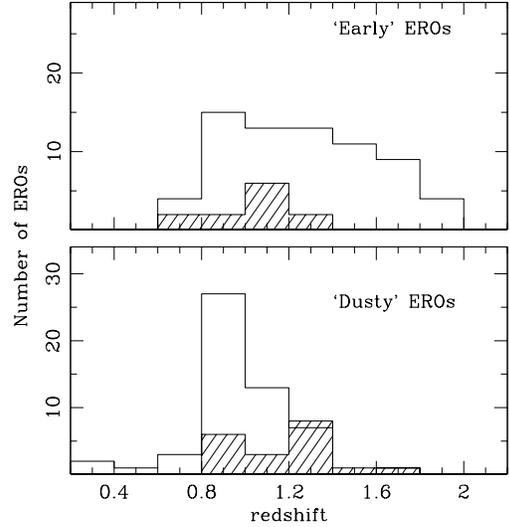,width=3in,angle=0}}
\caption
 {Open histograms are the photometric redshift distribution of the
 $K<19.5$\,mag  EROs in the PDS that are best fit with early (upper
 panel) or dusty (lower panel) template SEDs. For comparison we also
 plot the spectroscopic redshift distribution of $K<19.2$\,mag EROs
 identified in the {\sf K20} survey.
 }
\label{fig_distz}
\end{figure}

Figure \ref{fig_colour} presents the $I-K$ against $J-K$ colour-colour
plot introduced by Mannucci \& Pozzetti (2001) to discriminate between 
dusty active and evolved EROs. Only sources in our sample with
available $J$-band photometry (detection or upper limit) are plotted
here. There is reasonable agreement between the classification using
the SED template fitting method and the optical/NIR colours of our
sample. Figure  \ref{fig_distz} plots the redshift distribution of the
$K<19.5$\,mag EROs that are best-fit by early-type and dusty templates
separately. These are broadly consistent with the distributions
discussed by Daddi et al. (2001) and Roche et al. (2003). Also shown
in Figure  \ref{fig_distz} are the spectroscopic redshift distribution
of $K<19.2$\,mag  EROs from the {\sf K20} survey (Cimatti et
al. 2002). There is reasonable agreement with our photometric redshift
estimates, although we find more $z\ga1.3$ early type EROs. At these
redshifts however, Cimatti et al. (2002) discuss that instrumental
issues are likely to affect the identification of absorption-line
systems resulting in incompleteness. 

To compare the properties of $K<19.5$\,mag EROs with those of the full
$K$-band selected population, we also apply the above photometric
redshift estimation method to all $K<19.5$\,mag non-ERO galaxies 
(i.e. $I-K<4$\,mag; total of 1208). Spectroscopic redshift information
available for  the brighter of these systems (total of 28) is used to 
assess the accuracy of the results. The 1\,sigma rms uncertainty of
the quantity $\delta z=z_{phot}-z_{spec}$ is estimated 0.06.  

\section{Radio detected EROs}
A total of 95 radio sources brighter than $60\mu$Jy overlap with the 
$Ks$-band survey region. Using a matching radius of 2\,arcsec we find
that 17 of them are associated with EROs, 14 of which have
$K<19.5$\,mag. The fraction of $K<19.5$\,mag  EROs with radio
counterparts is $8\pm2$ per cent (14/177). This is a factor of 2 higher
than the fraction of non-EROs ($I-K<4$\,mag) at the same magnitude limit
with radio counterparts ($4\pm1$ per cent),
suggesting a higher fraction of dusty active systems among
the ERO population. The properties of the radio matched EROs are
presented in Table \ref{tab_radio}. About half of the radio emitting EROs
are best-fit by dusty SEDs (8/17) with the remaining half assigned
either early type templates (4/17) or no SED (5/17). Three of the
dusty sources  are assigned redshifts $z<0.8$ lower than that expected
for EROs (i.e. $z\approx1$). These are interesting objects in their
own right, since the k-correction does not  strongly contribute to the
red colors observed. They may resemble the ERO already studied in the
PDS, PDFJ011423 (Afonso et al. 2001), an extreme example of a dusty 
star-forming dominated ERO at relatively low redshift, $z=0.65$.   

Assuming that the radio emission is due to starburst activity and
adopting the relation between radio luminosity ($L_{\rm 1.4\, GHz}$)
and star-formation rate  (SFR) of Bell et al. (2003) we estimate SFRs
for stars in the mass range  $\rm 0.1-100\,M_{\odot}$ of about $10^2 -
10^3\, \rm M_{\odot} \, yr^{-1}$ for EROs at $z\approx1$. Also,
adopting the radio-FIR correlation (Helou, Soifer \& Rowan-Robinson
1985) we estimate FIR luminosities of about $10^{11} - 10^{12} \, \rm
L_{\odot}$ placing these systems in the class of Luminous and
Ultra-Luminous Infrared Galaxies (LIGs; ULIGs). 

\begin{table*}
\footnotesize
\begin{center}
\begin{tabular}{lcc ccc ccc ccc}
\hline
\# & $\alpha_{1.4}$ & $\delta_{1.4}$ & $\delta r$ & $I$ & $J$ & $K$ &
$S_{1.4}$ & $z$ & $\chi^2$ & $P_{1.4}$ & SFR \\
   & (J2000) & (J2000) & (arcsec) & (mag) & (mag) & (mag) & (mJy) & &
& (W/Hz) & ($\rm M_{\odot}/yr$)\\
\hline  
1 & 01 10 54.10 & $-$45 50 02.54 & 0.8 & $23.41\pm0.04$ & $-$ & $18.94\pm0.04$ & 0.070 & $1.40_{-0.04}^{+0.04}$ & 1.19 & 24.00 & $-$ \\
2 & 01 10 58.82 & $-$45 51 29.48 & 0.6 & $22.77\pm0.02$ & $-$ & $18.22\pm0.03$ & 0.100 & $0.92_{-0.06}^{+0.09}$ & 0.31 & 23.69 & 276 \\
3$^{\star}$ & 01 11 00.91 & $-$45 49 39.67 & 1.6 & $>$24.50 & $-$ & $19.77\pm0.06$ & 0.109 & $-$ & $-$ & $-$ & $-$ \\
4 & 01 11 02.63 & $-$45 51 34.96 & 0.1 & $21.67\pm0.01$ & $-$ & $17.59\pm0.02$ & 0.222 & $0.54_{-0.04}^{+0.02}$ & 1.00 & 23.45 & 158 \\
5 & 01 11 03.81 & $-$45 51 18.10 & 0.7 & $21.96\pm0.01$ & $-$ & $17.48\pm0.02$ & 0.086 & $0.62_{-0.06}^{+0.14}$ & 0.25 & 23.19 & 86 \\
6 & 01 11 09.66 & $-$45 48 19.43 & 1.1 & $23.47\pm0.03$ & $21.01\pm0.17$ & $19.13\pm0.04$ & 0.084 & $1.22_{-0.01}^{+0.05}$ & 0.65 & 23.93 & $-$ \\
7 & 01 11 13.13 & $-$45 51 23.11 & 0.8 & $23.32\pm0.03$ & $-$ & $18.83\pm0.04$ & 0.077 & $1.35_{-0.04}^{+0.06}$ & 0.35 & 24.01 & $-$ \\
8 & 01 11 14.21 & $-$45 50 03.21 & 0.7 & $22.36\pm0.02$ & $-$ & $18.01\pm0.02$ & 0.152 & $-$ & $>2.7$ & $-$ & $-$ \\
9 & 01 11 14.20 & $-$45 42 49.94 & 0.9 & $22.21\pm0.02$ & $19.88\pm0.07$ & $17.98\pm0.02$ & 0.086 & $0.25_{-0.01}^{+0.05}$ & 1.15 & 22.27 & 10 \\
10 & 01 11 15.23 & $-$45 41 00.51 & 0.3 & $22.92\pm0.02$ & $-$ & $18.28\pm0.03$ & 0.138 & $1.19_{-0.01}^{+0.08}$ & 2.61 & 24.12 & 735 \\
11 & 01 11 25.33 & $-$45 47 24.74 & 1.0 & $23.10\pm0.02$ & $21.35\pm0.21$ & $18.94\pm0.04$ & 0.148 & $0.98_{-0.24}^{+0.10}$ & 0.24 & 23.93 & 480 \\
12 & 01 11 36.53 & $-$45 41 54.74 & 1.2 & $22.12\pm0.01$ & $>$21.50 & $18.08\pm0.03$ & 0.101 & $-$ & $>2.7$ & $-$ & $-$ \\
13 & 01 11 37.00 & $-$45 40 36.98 & 1.9 & $23.05\pm0.02$ & $-$ & $18.99\pm0.04$ & 0.246 & $0.90_{-0.09}^{+0.08}$ & 0.33 & 24.06 & 641 \\
14$^{\star}$ & 01 11 58.39 & $-$45 40 28.28 & 1.6 & $>$24.50 & $-$ & $19.70\pm0.07$ & 0.089 & $-$ & $-$ & $-$ & $-$ \\
15 & 01 12 00.65 & $-$45 39 41.33 & 0.6 & $22.94\pm0.02$ & $-$ & $18.70\pm0.04$ & 0.138 & $0.82_{-0.03}^{+0.05}$ & 2.05 & 23.71 & $-$ \\
16 & 01 12 07.51 & $-$45 46 16.28 & 0.5 & $23.70\pm0.04$ & $-$ & $18.40\pm0.02$ & 0.101 & $1.13_{-0.02}^{+0.05}$ & 0.96 & 23.93 & 470 \\
17$^{\star}$ & 01 12 05.82 & $-$45 46 02.30 & 0.5 & $24.09\pm0.04$ & $-$ & $19.64\pm0.04$ & 0.070 & $-$ & $-$ & $-$ & $-$ \\
\hline
%\multicolumn{12}{l}{$^{\star}$No photometric redshift estimated because
%detected in less than 3 filters.}\\
%\multicolumn{12}{l}{1: ID number, 2, 3: right ascension and declination
%in J2000 of radio centroid, 4: radio/$K$-band position offset, 5, 6,
%7: $I$, $J$, $K$-band magnitudes}\\
%\multicolumn{12}{l}{8: radio flux density, 9, 10: photometric redshift
%and $\chi^2$ of the best fit. When $\chi^2>2.7$ no photo-z is listed ,
%11: radio power}\\ 
%  \multicolumn{12}{l}{12: SFR for those sources only that are best-fit
%by dusty SEDs.}
\end{tabular}

\begin{list}{}{}
\item $^{\star}$No photometric redshift estimated because
detected in less than 3 filters.
\item The columns are: 1: ID number; 2, 3: right ascension and declination 
in J2000 of radio centroid; 4: radio/$K$-band position offset; 5, 6,
7: $I$, $J$, $K$-band magnitudes; 8: radio flux density; 9, 10:
photometric redshift  and $\chi^2$ of the best fit. When $\chi^2>2.7$
no photo-z is listed; 11: radio power; 12: SFR for those sources only
that are best-fit by dusty SEDs.
\end{list}
\end{center}
\caption{The properties of EROs with radio counterparts.}
\label{tab_radio}
\normalsize
\end{table*}

\section{Radio stacking}\label{sec_stack}
For EROs that are not detected in the PDS radio survey we provide an
estimate of their mean radio properties using  the stacking analysis
method described by Hopkins et al. (2004) and Georgakakis et
al. (2005). We extract sub-images from the radio mosaic at
the location of the non-radio detected EROs, and construct  the
weighted average of the sub-images (weighted by 1/rms$^2$, to maximise
the resulting signal-to-noise, since the radio mosaic has a varying
noise  level over the image). Sub-images where low S/N emission
($>1.5\,\sigma$) is present at the location of the non-detected source
are excluded from the stacking, in order to avoid biasing the stacking
signal result by the presence of a small number of low S/N
sources. The stacking analysis above is applied separately to systems
that are best-fit by the dusty and early-type templates.
The results are presented in Table \ref{tab_stack}. For the
dusty EROs we estimate a mean radio flux density of $8.6\,\mu$Jy
significant at the $3\sigma$ confidence level ($1\sigma$ rms of $2.8\rm
\,\mu Jy$). This signal is likely to be associated with obscured
star-formation activity in the host galaxy, 
the stacked image has an rms noise of $2.7\,\mu$Jy, and a marginally
significant, $2.4\sigma$, detection at $6.4\,\mu$Jy. The radio
emission in these evolved systems may arise in low-level
AGN activity.

To test the sensitivity of these results to the dusty/early type 
classification scheme based on SED template fitting, we also
use the Pozzetti \& Mannucci (2001) $I-K$ vs $J-K$ plot to segregate
EROs into different types and repeat the stacking analysis. Only the
smaller subsample of EROs with $J$-band information available are
used. The results are also given in Table \ref{tab_stack}. We estimate
mean flux densities of $6.0$ and $7.5\,\mu$Jy for dusty and early EROs 
respectively, albeit with lower statistical significance, 1.6 and
$2.3\sigma$ respectively, because of the smaller number of sources
involved in the stacking. Nevertheless, these estimates are in
reasonable agreement, within the uncertainties, with those
found using the classification based on template fitting, suggesting
our results are robust.   

At the mean redshift of dusty and early type EROs estimated in 
section \ref{sec_photoz} the mean flux densities above (using the SED
fitting classification) translate to 1.4\,GHz luminosities of  $\rm
4.5 \times  10^{22}$ and $\rm 3.4 \times 10^{22}\, W \, Hz^{-1}$
respectively. We use a k-correction assuming a power law spectral
energy distribution of the form $S_{\nu} \propto \nu^{-\alpha}$ with
$\alpha=0.8$. In the case of evolved EROs, the observed mean radio
luminosity may originate in low-level AGN activity, found in
many ellipticals. Indeed, the stacking analysis provides sensitivities 
well below the radio-loud AGN limit ($\approx  10^{24} \rm W/Hz$;
Ledlow \& Owen 1996), allowing the detection of signal from lower
luminosity systems. For the the dusty  subsample the observed mean  
radio emission is likely associated with starburst activity (e.g.
LIGs; Sadler et al. 2002). Assuming this is the case, the mean
radio  luminosity of the dusty sub-population corresponds to an
average $\rm SFR(0.1-100\,M_{\odot}) \rm \approx 25 \, M_{\odot} \,
yr^{-1}$, adopting the calibration of Bell  (2003). This is of the
same order of magnitude as the mean ERO SFR estimate of Yan et
al. (2004), $\rm \approx 50 \, M_{\odot} \, yr^{-1}$,  based on $\rm
24\mu m$ Spitzer observations of the ELAIS-N1 region.  The factor of
two difference can be explained by the fact that Yan et al. (2004)
take the mean of the $\rm 24\mu m$ flux distribution of EROs detected
at these wavelength and convert that to SFR assuming $z=1$. If we
instead take the mode of their distribution, which is more
representative of the full dusty ERO  population and comparable to the
statistical analysis presented here, we find excellent agreement with
our results.   

\begin{table}
\footnotesize
\begin{center}
\begin{tabular}{lc cc}
\hline
sample & number     & $S_{1.4}$      &  significance \\
       & of sources & ($\rm \mu Jy$) &               \\
\hline
\multicolumn{4}{l}{SED fitting classification}\\
\hline
dusty  &   48       & 8.6 &  $3.0\sigma$\\
early  &   47       & 6.4 &  $2.6\sigma$\\
       &            &     &             \\
\hline
\multicolumn{4}{l}{$I-K$ vs $J-K$ classification}\\
\hline
%dusty  &   23       & 10.4 & $2.5\sigma$ \\
dusty  &   23       & 6.0 & $1.6\sigma$ \\
%early  &   26       &  9.2 & $3.2\sigma$ \\
early  &   26       &  7.5 & $2.3\sigma$ \\
\hline
\end{tabular}
\end{center}
\caption{Radio stacking results.}
\label{tab_stack}
\normalsize
\end{table}

\section{Star formation rate density}
In this section we use the radio emission of $Ks<19.5$\,mag  dusty
EROs  to explore their contribution to the global SFR density at
$z\approx1$. For those EROs that do not have a detected radio counterpart
we use the stacking analysis results of section \ref{sec_stack}. We
further assume that the radio flux density of the  dusty ERO
sub-population is dominated by star-formation rather than AGN
emission. This is a reasonable assumption since X-ray surveys suggest
that only 3-14 per cent of EROs have X-ray counterparts likely to be
associated with obscured AGN (Brusa et al. 2005 and references
therein). The upper limit in the X-ray/ERO identification rate 
range above occurs in the ultra-deep Chandra surveys (Alexander et
al. 2003) that are sufficiently sensitive to detect X-ray emission
from powerful starbursts at $z\approx1$. A fraction of the X-ray/ERO
associations in these fields are therefore, dominated by star-formation
rather than AGN activity. Moreover, only a small  fraction of the AGN
population ($\approx10$ per cent) show radio emission while, the radio
properties of some obscured AGNs are suggested to be dominated by
powerful starburst activity (e.g. Bauer 2002;  Georgakakis  et
al. 2004). The evidence above suggests that contamination of our
sample by AGN is likely to have little impact on the ERO
star-formation density estimation.  

We first determine the radio luminosity density using the photometric
redshift information to estimate the effective cosmological volume
probed by the $Ks$-selected sample  (e.g. $Ks<19.5$\,mag) using the
standard $1/V_{max}$ formalism  (e.g. Lilly et al. 1996; Mobasher et
al. 1999). The radio luminosity density is then converted to SFR
density using the calibration of Bell [2003;
$\rm SFR ( M = 0.1-100\,M_{\odot}) = L_{1.4} / 1.81\times10^{21}$, for
$L_{1.4}>6.4\times10^{21}$\,W\,Hz$^{-1}$]. For this  exercise we only
consider EROs that are best-fit by dusty template SEDs and have
photometric redshifts in the range $z=0.85-1.35$ (total of 41). We
estimate a radio luminosity density of $\rm (3.6 \pm 1.8) \times
10^{19} \, W \, Hz^{-1} \, Mpc^{-3}$ corresponding to $\rho_{\star}
\rm ( M = 0.1 - 100 \, M_{\odot}) = 0.02 \pm  0.01  \rm
M_{\odot}\,yr^{-1}\,Mpc^{-3}$. The dominant source of uncertainty in
this estimate is cosmic variance, which is assumed to be about 50
per cent of the ERO number counts (Somerville et al. 2004b). As
discussed in section \ref{sec_photoz} about 30 per cent of the
$K<19.5$\,mag EROs are excluded from the analysis because the either
have poor data (detected in less than 3 bands) or are not well fit
($\chi^2>2.7$) by the template SEDs used in this study. Assuming that
this population has similar redshift distribution and classification
mix (e.g. dusty vs early) with the $K<19.5$\,mag EROs used in our
analysis, the SFR density above should be  revised upward
by about 30 per cent. In the analysis that follows we do not take into
account this moderate incompleteness correction.  

\begin{figure}
 \centerline{\psfig{figure=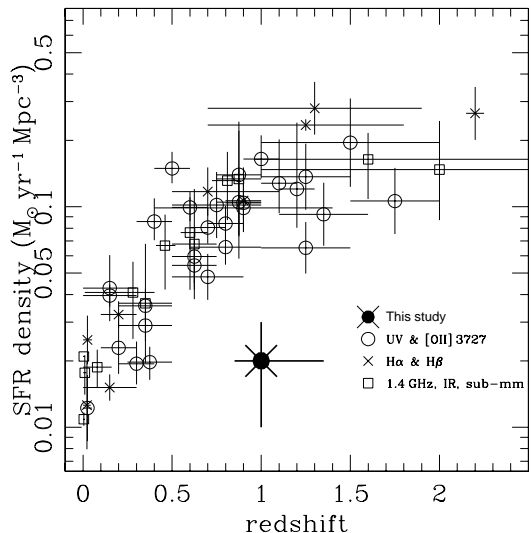,width=3in,angle=0}}
\caption
 {Global SFR density density as a function of redshift. The
 filled crossed circle is the $K<19.5$\,mag EROs contribution
 estimated in this paper plotted at the median redshift of the
 sample, $z=1$. This is compared with global SFR density compilation
 of Hopkins (2004). Points are coded by rest frame wavelength used to
 estimate the SFR: open circles: UV; crosses: $\rm H\alpha$ and $\rm
 H\beta$; 1.4\,GHz, IR or sub-mm: open squares.
 }\label{fig_sfr}
\end{figure}

Our results are plotted in Figure \ref{fig_sfr} at the median redshift
of $\approx1$ in comparison with the compilation of SFR densities from 
different wavelengths of Hopkins (2004). The $K<19.5$\,mag dusty ERO
estimate accounts for about 10 per cent of the total SFR density
estimated at $z\approx1$ ($\approx 0.15 \rm M_{\odot} \, yr^{-1} \,
Mpc^{-3}$). The fraction above should be considered a lower limit
however, since it   does not take into account EROs fainter than
$Ks=19.5$\,mag. This suggests that dust enshrouded starburst activity
in EROs is potentially a non-negligible component of the global SFR
at these redshifts.   

Cimatti et al. (2002) used the {\sf K20}  sample with a magnitude
cutoff $K<19.2$\,mag, similar to  the present study, and found
$\rho_{\star} \approx \rm 0.015 \, M_{\odot} \, yr^{-1} \, Mpc^{-3}$
in the range $z=0.85-1.3$, in fair agreement with our result. Unlike
our study however, these authors used the $ \rm [OII]\,3727\AA$ line
as a SFR estimator and applied a mean dust correction of
$E(B-V)\approx 0.5$, which was found to be consistent with the average
UV/optical continuum of emission-line EROs in their sample. The
agreement between our dust-independent estimate and that of Cimatti
et al. (2002), after taking into account reddening, suggests that a
mean dust extinction of at least $E(B-V)\approx 0.5$ is required to
provide an unbiased view of the ERO star-forming population. It is
interesting to note that recent hydrodynamical  simulations by
Nagamine et al. (2005) also invoke a uniform extinction of
$E(B-V)\approx 0.4$ for their model galaxies to explain the red
observed colours of EROs.   

Smail et al. (2002) used deep radio data (1.4\,GHz) to estimate
the SFR density of $K<20.5$\,mag  EROs in the range $z=0.8-1.5$. We
consider their dusty ERO  sub-sample (total of 20) classified on 
the basis of SED template fitting. This is directly comparable to the
ERO sample used here to estimate the SFR density. Using the
$L_{1.4}$-to-SFR conversion adopted here, we estimate a SFR density of      
$\rho_{\star} \rm ( M = 0.1 - 100 \, M_{\odot}) \approx \rm 0.04 \,
M_{\odot} \, yr^{-1} \, Mpc^{-3}$ for the Smail et al. (2002) $K
\approx 20.5$\,mag dusty EROs. This factor of 2 difference compared to
our estimate suggests that EROs fainter than our $K=19.5$\,mag limit
make a large contribution to the SFR density at $z \approx 1$, and the
ERO component may in fact  be a large contribution (about 25 per cent)
to the SFR density at this redshift. Indeed, Smail et al. argue that
the observed break in the number counts of EROs at $K\approx19.5$
(McCarthy et al. 2001; Smith et al. 2002) is because passive EROs make
a sizable contribution to the counts at bright magnitudes ($K \la
20$\,mag) while dusty active EROs make up the bulk of the population
at fainter limits. We also note that incompleteness corrections in the
Smail et al. (2002) SFR density (e.g. sources that are not
well fit by their template SEDs) would revise their result upward by
about 45 per cent increasing it to $\rho_{\star} \rm ( M = 0.1 - 100
\, M_{\odot}) \approx \rm 0.06 \, M_{\odot} \, yr^{-1} \, Mpc^{-3}$
(using the $L_{1.4}$-to-SFR conversion adopted here). This
incompleteness correction is comparable to that estimated for the PDS
EROs.

In the next section we will present evidence that the $K<19.5$ ERO
sample and the dusty sub-population in particular, are likely to be
complete to the stellar mass limit $\rm M \approx 5 \times 10^{10} \,
M_{\odot}$. This combined with the above results suggests that about
half of the dusty ERO SFR density at $z\approx1$ arises in systems
with mass $\rm \ga 5 \times 10^{10} \, M_{\odot}$ (i.e. those detected
in this study) with the remaining half in less massive galaxies fainter
than $K\approx19.5$\,mag. This is in contrast to the local Universe
where the most massive galaxies have little, if any, on-going
star-formation activity (e.g. Kauffmann et al. 2004). This trend
continues to $z\approx1$ although at these redshifts numerous studies
also find evidence for a population of massive galaxies that
experience starburst activity in agreement with our result (e.g. Cowie
et al. 1996; Drory et al. 2004; Fontana  et al. 2004).   

%Elliptical galaxies are proposed as the most likely descendants of
%EROs. The median luminosity of the dusty subsample studied here is
%$M_K=-25.1$. Assuming passive evolution from $z=1$ to the present day,
%this luminosity corresponds to $M_K\approx24.4$\,mag at $z=0$,
%i.e. similar to the characteristic luminosity of local ellipticals
%$M_K^{*}\approx24.2$\,mag (Kochanek et  al. 2001). Therefore, it is
%possible that both the early-type and many of the dusty EROs in our
%sample may evolve into $L>L^{*}$ present-day ellipticals.      

\section{Stellar mass density}
Next we estimate the stellar mass density of EROs at $z\approx1$ in
comparison with that of the full sample of  $K$-band selected
galaxies with $K<19.5$. Galaxy  masses are estimated  using $K$-band
luminosities and the mass--to--light ratio, $M/L_K$, of the best fit
SED for each system (e.g. Caputi et al. 2005). These are in fair
agreement with the mean $M/L_K$ ratios of NIR selected galaxies at
$z=1.0-1.5$ estimated by Fontana et al. (2004) from the {\sf K20}
sample ({\sf http://www.arcetri.astro.it/$\sim$k20}). These authors
estimate   stellar masses by fitting a range of template SEDs to
multiwaveband photometry ($UBVRIzJKs$) of spectroscopically identified
$K<20$\,mag galaxies. They adopt a grid of SEDs with exponentially
declining star-formation rates, a Salpeter IMF and an SMC extinction
law to estimate $M/L_K$ ratios in the redshift range $0 \la z \la 2$. 

The estimated masses, particularly for early-type EROs, are likely to
be representative of the population. This is because compared to
shorted wavelengths the $K$-band is more closely associated to the
integrated galaxy stellar mass, is less affected by bursts of
star-formation, dust extinction, and the galaxy type, while there is
only weak dependence of the $M/L_K$ on redshift. We caution the reader
however, that dusty EROs are more difficult to model and to estimate
mean $M/L_K$  ratios for, since they are likely to have mixed stellar
populations, large extinction and complex dust covering factors. 

We first estimate the number density of EROs in comparison with those
of the full $K<19.5$\,mag sample using the standard $1/V_{max}$
method. We note that the magnitude limit $K=19.5$ corresponds to a
mass of  $\rm M \ga 5\times 10^{10} \, M_{\odot}$ at $z=1$, adopting
the mean k-corrections and mass-to-light ratios used here ($ M / L_K
\approx 0.6$). Our sample is therefore, nearly complete for galaxies
more massive than $ \rm \ga 5\times 10^{10} \, M_{\odot}$. For the
redshift range $0.85-1.35$ and $\rm M> 5\times 10^{10} \, M_{\odot}$ we
estimate number densities of $\rm (4.7 \pm 2.4 ) \times 10^{-4} \,
Mpc^{-3}$ for EROs and $\rm (10.9 \pm 3.1 ) \times 10^{-4} \, Mpc^{-3}$
for the full $K<19.5$\,mag  galaxy population. The dominant source of
uncertainty in these estimates is cosmic variance, which is assumed to
be of about 50 and 30 per cent of EROs and non-EROs respectively
(Somerville et al. 2004b). These densities are
in good  agreement with the results of Caputi et al. (2005) at similar
redshifts. As also noted by these authors, EROs are a sizable component of
galaxies more massive than $M> 5\times 10^{10} \, M_{\odot}$ at
$z\approx1$, representing about 50 per cent of the
population in the PDS. Moreover, both dusty and early-type EROs
contribute almost equally to this fraction with number densities of
about $\rm 2.7 \times 10^{-4}$ and  $\rm 2 \times 10^{-4} \,
Mpc^{-3}$. This suggests that both the dusty active and the early-type
EROs at $z\approx1$ have already assembled a sizable fraction of their
stellar mass.  

The evolution of EROs to $z=0$ remains uncertain although elliptical
galaxies are proposed as their descendants. Under this assumption we
explore connections between the NIR properties of the two
populations.  The median luminosity of the PDS EROs is estimated 
$M_K=-25.1$\,mag. Assuming passive evolution from $z=1$ to the present 
day this luminosity corresponds to $M_K \approx -24.4$\,mag at 
$z=0$, i.e. similar to the characteristic absolute magnitude of local
ellipticals $M_K^{*}\approx-24.2$\,mag (Kochanek et
al. 2001). The evidence above suggests that if EROs evolve into nearby
ellipticals they occupy the bright end ($L \ga L^{*}$)  of the luminosity
function of these systems. 

\begin{figure}
 \centerline{\psfig{figure=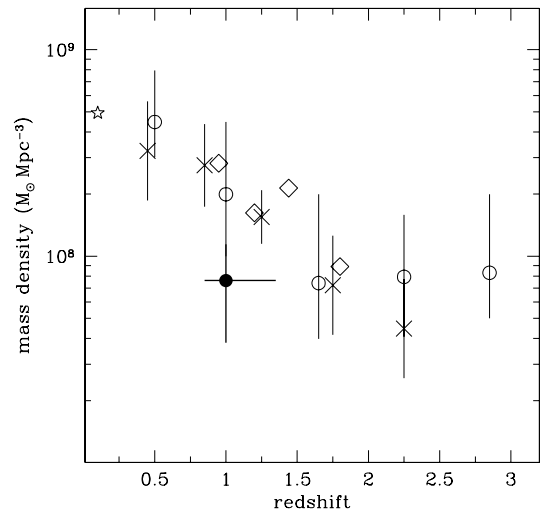,width=3in,angle=0}}
\caption
 {Global mass density as a function of redshift. The filled circle is
 the $K<19.5$\,mag ERO contribution estimated in this paper plotted
 at the median redshift of the sample, $z=1$. Star: Cole et al. (2000);
 crosses: Fontana et al. (2004); open circles: Fontana et al. (2003);
 diamonds: Glazebrook et al. (2004).  
 }
\label{fig_mass}
\end{figure}

The mass density of early and dusty EROs at $z=0.85-1.35$ are
presented in Table \ref{tab_res} and are plotted in Figure
\ref{fig_mass}. The errorbars are estimated assuming 50 per cent
cosmic variance to the ERO number counts. Dusty and early type systems
contribute almost equally to the ERO mass assembly at this redshift
range. Figure \ref{fig_mass} compares our results with estimates of
the global mass density out to $z\approx3$. Although there is some
uncertainty in the determination of the total stellar mass density at
$z\approx1$, adopting a value of $\rm \approx2.5\times10^8  \,
M_{\odot} \, Mpc^3$ we find that  the $K<19.5$ EROs represent about 30
per cent of that,  suggesting that they are a non-negligible component
of the mass assembly at this redshift. This fraction is likely to
represent a lower limit since EROs fainter than the $Ks=19.5$ are not
taken into account. We address this point by comparing with the
stellar mass density of all  $K<19.5$\,mag galaxies in the range
$0.85<z<1.35$. Any incompleteness biases are likely to affect both this
and the EROs samples in almost the same way. We estimate a stellar
mass density of $\rho_m= \rm  (1.7\pm0.5)  \times 10^{8} \, M_{\odot}
\, Mpc^{-3}$ for $K<19.5$\,mag galaxies. Therefore the  EROs studied
here represent about 50 per cent of the stellar mass density  of the
full  $Ks<19.5$ sample. Although not {\it all} massive galaxies at
$z\approx1$ are EROs the evidence above underlines the significance of
this population with regards to the mass assembly of the Universe at
$z\approx1$.

\begin{table}
\footnotesize
\begin{center}
\begin{tabular}{lc cc }
\hline
sample & number     & $\rho_{m} ( \rm >5\times10^{10}\, M_{\odot})$ &
$\rho_{\star} (\rm 0.1-100\,M_{\odot})$ \\
       & of sources & ($\rm M_{\odot}\,Mpc^{-3}$) &
($\rm M_{\odot}\,yr^{-1}\,Mpc^{-3}$) \\
\hline
dusty  &   41       & $4.1 \pm 2.0 \times 10^7$ &  $0.02\pm0.01$ \\
early  &   34       & $3.5 \pm 1.7 \times 10^7$ &       $ - $    \\
all    &   75       & $7.6 \pm 3.2 \times 10^7$ &       $ - $   \\
\hline
\end{tabular}
\end{center}
\caption{Mass and SFR density for $K<19.5$ EROs in the redshift range $0.85<z<1.35$.}
\label{tab_res}
\normalsize
\end{table}

\section{Summary \& conclusions}
In this paper we use an $\approx180 \rm \, arcmin^2$ $Ks$-band survey
overlapping with ultra-deep radio observations and  multiwaveband 
photometry ($UBVRIJ$) to estimate the SFRs and the stellar
masses of EROs with $K<19.5$\,mag. Template SEDs are fit to the
broad-band optical/NIR photometric data to discriminate between
dusty and early-type EROs and to  determine their photometric
redshifts. This classification is found to be in fair agreement with
methods using optical/NIR colours (e.g. $I-K$ vs $J-K$).

About 8 per cent of the $K \la 19.5$\,mag EROs have radio
counterparts to the flux density limit of about $60\rm \, \mu Jy$. For
the remaining sources we use radio stacking analysis to constrain
their mean radio properties. We estimate a stacked signal of about
$9$ ($3\sigma$) and $6 \, \rm \mu Jy$ ($2.5\sigma$) for dusty and
early-type EROs respectively. Assuming that the radio emission in the
dusty sub-population is due to star-formation activity we estimate a
mean star-formation rate of $\rm SFR  =  25\,M_{\odot}
\, yr^{-1}$ at $z=1$ free from dust obscuration effects. The radio
detected EROs at $z\approx1$ have, on average, much higher SFRs in the
range $\rm 100-1000 \, M_{\odot} \, yr^{-1}$. Combining these results
with the photometric redshift estimates we find that the SFR density
of the Universe in the range $0.85-1.35$ due to $K < 19.5$\,mag EROs
is $\rm 0.02\pm0.01 \, M_{\odot} \, yr^{-1} \, Mpc^{-3}$. This should
be considered a lower limit since EROs fainter than $K = 19.5$\,mag
are not taken into account. Comparison with deeper samples suggests
that correcting for these biases will likely
increase the estimate above by at least a factor of 2. We also argue
that the systems responsible for the observed star-formation density
are dusty starbursts more massive than  $\rm M = 5 \times  10^{10} \,
M_{\odot}$. Less massive dusty EROs lie below the magnitude limit
$K=19.5$\,mag at $z\approx1$ and are the systems responsible for the
missing (at least factor of 2) SFR density at $z\approx1$. Comparison
of the dust-independent SFR density estimated here with that of
similarly selected ERO samples using optical emission lines, suffering
dust attenuation, suggests a mean dust reddening of at least 
$E(B-V)\approx0.5$ for this population.   

We further use mass-to-light ratios of the best-fit template SED to
convert the $Ks$-band luminosity of EROs to stellar mass. We find that
EROs contribute about 50 per cent to the total number density of
galaxies with stellar mass $\rm M >  5 \times  10^{10} \,  M_{\odot}$
at $z\approx1$. This fraction is almost equally split between dusty and
early type systems. We further estimate that the $K<19.5$ EROs
represent about 50 per cent of the global mass density in the redshift
range $z=0.85-1.35$, after taking into account incompleteness due to
the magnitude limit $K=19.5$\,mag. This indicates  that these systems
are also a non-negligible component of the Universe mass build-up at
these redshifts. The ERO mass density above is also almost
equally split between the dusty and early type subpopulations. 

\section{Acknowledgments}
We thank the anonymous referee for useful comments and suggestions
that improved this publication. Part of the data presented in this
paper are available at 
{\sf http://www.atnf.csiro.au/people/ahopkins/phoenix/}. This work is 
based on observations collected at the European Southern Observatory, 
Chile, ESO 66.A-0193(A). JA gratefully acknowledges the support from 
the Science and Technology Foundation (FCT, Portugal) through the
research grant POCTI-FNU-43805-2001.

\end{document}